\documentclass[%
reprint,
 amsmath,amssymb,
 aps,
superscriptaddress,
nofootinbib,
floatfix
]{revtex4-2}

\usepackage{lipsum}
\usepackage{tikz}
\usepackage{graphicx}
\usepackage{dcolumn}
\usepackage{bm}
\usepackage{siunitx} 
\usepackage{hyperref}
\usepackage[mathlines]{lineno}
\usepackage{chemformula}
\usepackage{sidecap}
\usepackage{multirow}
\sidecaptionvpos{figure}{c}
\usepackage[capitalise]{cleveref}
\Crefname{figure}{Fig.}{Figs.}
\Crefname{table}{Tab.}{Tabs.}
\Crefname{equation}{Eq.}{Eqs.}

\usepackage{xspace}
\usepackage{colortbl}
\usepackage{ifthen}

\newcommand{\bvk}[1]{\ifthenelse{\boolean{show_comments}}{\textcolor{orange}{[{\bf BvK}: #1]}}{}}
\newcommand{\sk}[1]{\ifthenelse{\boolean{show_comments}}{\textcolor{brown}{[{\bf SK}: #1]}}{}}
\newcommand{\mk}[1]{\ifthenelse{\boolean{show_comments}}{\textcolor{teal}{[{\bf MK}: #1]}}{}}
\newcommand{\ft}[1]{\ifthenelse{\boolean{show_comments}}{\textcolor{purple}{[{\bf FT}: #1]}}{}}
\newcommand{\tf}[1]{\ifthenelse{\boolean{show_comments}}{\textcolor{violet}{[{\bf TF}: #1]}}{}}
\newcommand{\bem}[1]{\ifthenelse{\boolean{show_comments}}{\textcolor{magenta}{[{\bf BM}: #1]}}{}}

\newcommand{\kalpha}[0]{$\mathrm{K}_\alpha$}
\newcommand{\kbeta}[0]{$\mathrm{K}_\beta$}
\newcommand{\fe}[0]{$^{55}\mathrm{Fe}$}
\newcommand{\mn}[0]{$^{55}\mathrm{Mn}$}

\def \delight {DELight\xspace}

\begin{document}

\title{Optimum filter-based analysis for the characterization of a high-resolution magnetic microcalorimeter towards the DELight experiment}


\author{Francesco Toschi}\email{francesco.toschi@kit.edu}
\affiliation{Institute for Astroparticle Physics (IAP), Karlsruhe Institute of Technology (KIT), 76131 Karlsruhe, Germany}
\affiliation{Kirchhoff-Institute for Physics, Heidelberg University, Im Neuenheimer Feld 227, 69120 Heidelberg, Germany}

\author{Benedikt Maier}
\affiliation{Institute of Experimental Particle Physics (ETP),  Karlsruhe Institute of Technology (KIT), 76131 Karlsruhe, Germany}

\author{Greta Heine}
\affiliation{Institute of Experimental Particle Physics (ETP),  Karlsruhe Institute of Technology (KIT), 76131 Karlsruhe, Germany}

\author{Torben Ferber}
\affiliation{Institute of Experimental Particle Physics (ETP),  Karlsruhe Institute of Technology (KIT), 76131 Karlsruhe, Germany}

\author{Sebastian Kempf}
\affiliation{Institute of Micro- and Nanoelectronic Systems (IMS), Karlsruhe Institute of Technology (KIT), 76131 Karlsruhe, Germany}
\affiliation{Institute for Data Processing and Electronics (IPE), Karlsruhe Institute of Technology (KIT), 76131 Karlsruhe, Germany}

\author{Markus Klute}
\affiliation{Institute of Experimental Particle Physics (ETP),  Karlsruhe Institute of Technology (KIT), 76131 Karlsruhe, Germany}

\author{Belina von Krosigk}%
\affiliation{Kirchhoff-Institute for Physics, Heidelberg University, Im Neuenheimer Feld 227, 69120 Heidelberg, Germany}
\affiliation{Institute for Astroparticle Physics (IAP), Karlsruhe Institute of Technology (KIT), 76131 Karlsruhe, Germany}

\date{\today}

\begin{abstract}

Ultra-sensitive cryogenic calorimeters have become a favored technology with widespread application where eV-scale energy resolutions are needed. 
In this article, we characterize the performance of an X-ray magnetic microcalorimeter (MMC) using a \fe~source.
Employing an optimum filter-based amplitude estimation and energy reconstruction, we demonstrate that a FWHM resolution of $\Delta E_\mathrm{FWHM} = \left(\num{1.25(17)}\text{\scriptsize{(stat)}}^{+0.05}_{-0.07}\text{\scriptsize{(syst)}}\right)\unit{\electronvolt}$ can be achieved, leading to an unprecedented energy resolving power $E/\Delta E_\mathrm{FWHM}\sim{4700}$ among existing energy-dispersive detectors for soft and tender X-rays.
We also derive the best possible resolution and discuss limiting factors affecting the measurement.
The analysis pipeline for the MMC data developed in this paper is furthermore an important step for the realization of the proposed superfluid helium-based experiment \delight, which will search for direct interaction of dark matter particles with masses below \SI{100}{\mega\electronvolt\per\text{\ensuremath{c}}\squared}.
\newline
\newline

\end{abstract}

\maketitle

\section{\label{sec:intro}Introduction}

Over the past decade, the field of cryogenic calorimetry has significantly evolved, yielding detectors with eV-scale energy thresholds and resolutions \cite{Kempf:2018,tes_solar_resolution,10.1063/1.4936793,10.1063/1.4726279,Ren:2020gaq}.
These detectors have a wide range of applications across various scientific fields due to their exceptional sensitivity to minute energy depositions.
Some of their key applications include X-, gamma-ray and mass spectroscopy~\cite{Herdrich_2023,PRETZL2000114,10.1063/1.4930036}, quantum information processing~\cite{Sauvageot_2020}, neutrino research~\cite{Gastaldo:2013wha,Nucciotti:2018vyc}, and searches for dark matter (DM) particles.
In particular, the low energy threshold of cryogenic detectors allows the investigation of DM parameter space which has been inaccessible in the past, i.e., the parameter space of light dark matter (LDM) candidates with masses notably below the typical GeV- to TeV-scale of weakly interacting massive particles (WIMPs)~\cite{Schumann:2019eaa}.
A breakthrough in the energy resolution of cryogenic detectors was achieved by the SuperCDMS collaboration in 2018, which reached a resolution of \SI{11}{\eV} employing transition edge sensors (TES)~\cite{SuperCDMS:2018mne}.
Nowadays, the best resolution for a cryogenic detector-based experiment is reached by the CRESST collaboration, with a baseline energy resolution of around \SI{1.4}{\eV} employing TES~\cite{CRESST:2022lqw}.

Among the many cryogenic detectors currently available, magnetic microcalorimeters (MMCs) are a particularly promising technology as they provide not only an outstanding energy resolution, but also a large dynamic range and an excellent linearity \cite{Kempf:2018}.
This work presents the analysis based on the optimum filter (OF) of the calibration data acquired using the MMC discussed in~\cite{MMC_paper}, which was optimized for the detection of soft and tender X-rays.
The presented analysis improves the previous full-width half-maximum (FWHM) resolution of \SI{1.8}{\eV} obtained using a template fitting~\cite{krantz_thesis} to $\Delta E_\mathrm{FWHM} = \left(\num{1.25(17)}\text{\scriptsize{(stat)}}^{+0.05}_{-0.07}\text{\scriptsize{(syst)}}\right)\unit{\electronvolt}$ for X-rays of energy \SI{5.9}{\keV}.
The MMC remarkable performance sets it apart as the top-performing energy-dispersive cryogenic detector, despite the limitation coming from the thermal fluctuations of the experimental setup.

Since MMCs are also particularly suited for the development of large-area detectors, the future superfluid $^4\mathrm{He}$-based \delight experiment for direct search for LDM will deploy this technology~\cite{vonKrosigk:2022vnf}.
Although the expected resolution will be worse than the one achieved for these X-ray microcalorimeters because of the larger absorber area, the analysis and processing framework developed for this X-ray MMC lays the foundation for future data processing and analysis in the context of \delight.

This paper is organized as follows.
The MMC detector is briefly described in \cref{sec:mmc}, followed by an overview of the experimental setup and the detector operation in \cref{sec:operation}.
\cref{sec:analysis} details the analysis of the MMC data leading to the final results.
The main findings are summarized in \cref{sec:conclusion}.

\section{\label{sec:mmc}Magnetic microcalorimeter}

The operation of an MMC involves the conversion of an energy deposition $\delta E$ into a variation in the magnetic flux.
The interaction primarily occurs in an absorber which is thermally coupled to a paramagnetic temperature sensor that is situated in a weak external magnetic field.
Upon energy deposition, the temperature of the sensor increases by $\delta T = \delta E/C_\mathrm{tot}$, where $C_\mathrm{tot}$ is the total heat capacity of the system.
According to Curie's law, this leads to a change in magnetization $\delta M \propto \partial M/\partial T \cdot \delta T$, which causes a variation of the magnetic flux $\delta \Phi \propto \delta M$.
A direct current superconducting quantum interference device (dc-SQUID) converts this small change in magnetic flux $\delta \Phi$ into a voltage change.

The data used in this work were acquired using the second variant MMC prototype as described by M.~Krantz \emph{et al.} in \cite{MMC_paper}.
This MMC integrates the Ag:Er paramagnetic temperature sensor directly into the dc-SQUID loop to enhance the magnetic flux coupling and minimize the influence of SQUID noise on the achievable energy resolution~\cite{Kempf:2018}.
The SQUID consists of two meander-shaped coils, which are connected in parallel.
One of these superconducting coils forms the SQUID loop, interrupted by two Josephson tunnel junctions, while the other coil below it generates the necessary bias magnetic field.
The absorber is made of electroplated gold with dimensions \qtyproduct{150 x 150 x 3}{\um}, resulting in a total heat capacity of \SI[per-mode = symbol]{0.1}{\pico\joule\per\K} at \SI{20}{\milli\kelvin} and providing absorption probabilities of \SI{98}{\percent} for X-rays below \SI{5}{\keV} and around \SI{50}{\percent} up to \SI{10}{\keV}.
To minimize athermal phonon escape into the substrate, the absorber is designed in a ``tetrapod geometry'', extending over the sensor.
This increases the path available for phonons to thermalize, while reducing the volume with a direct line of sight to the sensor.

\section{\label{sec:operation}Experimental setup and operation}

\begin{figure}[t]
\centering
\includegraphics[width=\columnwidth]{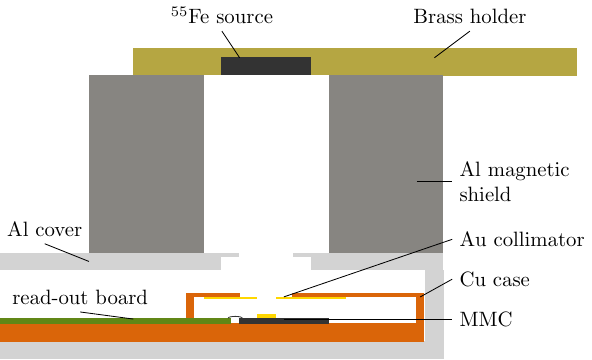}
\caption{Schematic view of the experimental setup for the calibration of the MMC using a \fe~source.}
\label{fig:setup}
\end{figure}

The setup used to characterize the MMC was directly mounted to the mixing chamber platform of a $\ch{^3He}/\ch{^4He}$ dilution refrigerator which was operated at its base temperature of about \SI{7}{\milli\kelvin}, without any active stabilization.
Due to the SQUID Joule heating, the sensor operated at a temperature of around \SI{18}{\milli\kelvin}, as discussed in \cite{MMC_paper}.
Assuming that the thermal fluctuations of the MMC followed the fluctuations of the refrigerator, the temperature stability could be verified within $\Delta T/T\simeq\num{6.4e-4}$, which is the measurement uncertainty of the refrigerator thermometer.
The MMC and its readout board were mounted on a copper case enclosed by an aluminum cover that works as a superconducting magnetic shield at the operational temperature.
The Al cover features a small slit which allows X-rays from a nearby calibration source to reach the detector.
As the presence of this opening hinders the shielding potential of the cover, an additional long Al shield is placed above it to reduce the leakage of external magnetic fields.
The slit is aligned with a \SI{15}{\um} thick gold collimator placed right above the MMC which avoids X-rays to hit the detector substrate, which would lead to temperature fluctuations and spurious signals.
The {\fe } X-ray source is mounted on a brass holder and placed right above the extended Al shield, around \SI{5}{\cm} from the MMC.
A schematic cross section view of the setup is shown in \cref{fig:setup}.

The iron source is contained in a stainless steel case with a \SI{200}{\um} thick beryllium window.
The {\fe } isotope undergoes electron capture into {\mn }, leaving a vacancy typically in the K-shell that is filled by an electron from a higher energetic shell, producing a cascade of X-rays and/or Auger electrons.
Due to its low energy, the emitted radiation coming from de-excitation into non-K shells is blocked by the beryllium window.
The highly energetic X-rays that reach the MMC arise from the K-shell vacancy being filled either by an L-shell electron, known as {\kalpha } transition, or by an M-shell electron, {\kbeta } transition.
The {\kalpha } spectral line exhibits fine structure splitting into two lines: $\mathrm{K}_{\alpha1}$ with an energy of \SI{5.899}{\keV} and natural width \SI{2.47}{\eV}, and $\mathrm{K}_{\alpha2}$ with an energy of \SI{5.888}{\keV} and natural width \SI{2.92}{\eV}~\cite{PhysRevA.56.4554}.
The \kbeta, although affected as well by the fine structure $LS$-coupling, does not present a similar splitting.
Its X-ray energy distribution peaks at \SI{6.49}{\keV} with a natural width of \SI{2.97}{\eV}.
The energy spectrum of these characteristic X-rays is further described in \cref{sec:spectrum}.

\begin{figure}[ht]
\centering
\includegraphics[width=\columnwidth]{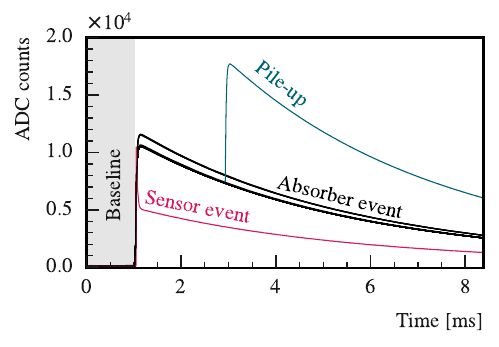}
\caption{\label{fig:ex_traces} Small subset of 50 traces from the \fe~calibration run. A total of 48 overlapping black traces are from X-rays interacting in the absorber and they are considered good events. The remaining two traces correspond to X-rays interacting directly in the paramagnetic sensor (magenta) and pile-up of events in the absorber (dark blue). The first 4000 samples (gray selection) are used to determine the trace baseline.}
\end{figure}

The {\fe } source was used to calibrate the MMC described in \cref{sec:mmc}.
The voltage response of its coupled dc-SQUID was read out by a custom-made multichannel data acquisition (DAQ) system discussed in~\cite{Mantegazzini_2021}.
It provided a maximum sampling rate of \SI{125}{\mega\hertz} with a 16-bit voltage resolution and an independent trigger for each channel.
The acquisition was triggered using a constant fraction discriminator, as the signals are expected to have different amplitudes but same rise time.
The high sampling rate of the DAQ allowed for a precise trigger time, while the acquired data, known as traces, were stored with a downsampling factor of 32 to reduce the necessary disk space.
With a total length of 32768 samples, a single trace has a duration of about \SI{8.4}{\ms} with the triggering time slightly after \SI{1}{\ms}.
For this reason the baseline of each trace is obtained from the first 4000 samples (around \SI{1}{\ms}).
The calibration lasted around \SI{50}{\min} with a trigger rate of \SI{2.94(3)}{\hertz}: a subset of acquired traces is shown in \cref{fig:ex_traces}.

\section{\label{sec:analysis} Data analysis}
\subsection{\label{sec:selection} Event selection}
A set of essential selection criteria is applied to the data in order to discard pile-up events, traces acquired during unstable conditions and events happening directly in the sensor or its substrate (see \cref{fig:ex_traces}).
These consist of selections of the signal rise-time, the mean value of the entire trace, and a loose selection on the chi-square value ($\chi^2$) from fitting a simple template to the trace in the time domain.
Additionally, to ensure detector stability, further criteria are applied to the standard deviation of the baseline and its offset, which comes from a linear fit.
The acceptance of these selection criteria is \SI{57(1)}{\percent}.
The level of the baseline is an indicator for the temperature of the detector right before an X-ray photon hits the sensor.
To study the effect of temperature variations, we therefore consider an additional selection on the mean value of the baseline, together with the aforementioned selections on its standard deviation and offset.
When applied, we refer to the dataset as ``tight selection'' with an additional acceptance of \SI{74(3)}{\percent} and characterized by a more stable detector temperature.
In case the selection is not applied, we refer to it as ``loose selection''.

\subsection{\label{sec:filter}Optimum filtering and energy reconstruction}
To achieve the highest possible resolution, the energy is reconstructed using an estimator based on the optimum filter (OF)~\cite{Gatti:1986cw,Wilson:2022quj}.
This consists of a finite-impulse-response filter which minimizes a $\chi^2$ value in the frequency domain, $\nu$.
This value is defined as
\begin{equation}\label{eq:of_chi2}
    \chi^2=\sum_\nu\frac{\left[S(\nu) - a A(\nu)\right]^2}{J(\nu)},
\end{equation}
where $S(\nu)$ is the signal, $A(\nu)$ is the finite-size signal template, and $J(\nu)$ is the power spectrum density (PSD), all in the frequency domain.
The amplitude obtained by minimizing $\chi^2$ is
\begin{equation}\label{eq:of_amp}
    a = \frac{\sum_\nu S^\ast(\nu) A(\nu)/J(\nu)}{\sum_\nu |A(\nu)|^2/J(\nu)}.
\end{equation}
The employed OF allows for the time shift of the signal with respect to the template, possible by considering a time delay $t_0$ in the time domain signal $A(t-t_0)$, which translates into a phase in frequency domain $e^{-2\pi it_0\nu}A(\nu)$.
The phase-shifted template is included in ~\cref{eq:of_amp} and the delay $t_0$ that maximizes $a$ is chosen, as it is possible to show that it also minimizes $\chi^2$~\cite{Filippini_thesis}.

\begin{figure}[b]
\centering
\includegraphics[width=\columnwidth]{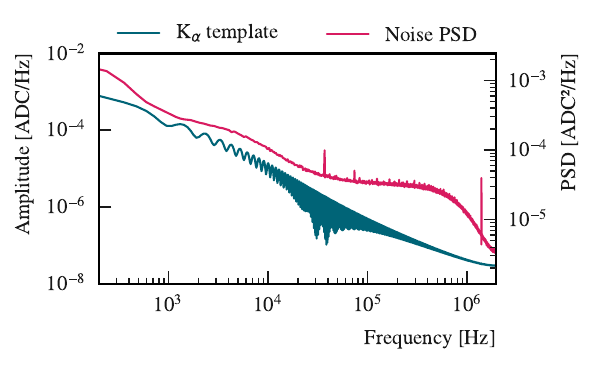}
\caption{\label{fig:psd} Amplitude of the template from the {\kalpha } energy region (dark blue, left y-axis) and noise PSD (magenta, right y-axis).}
\end{figure}

\begin{figure}[t]
\centering
\includegraphics[width=\columnwidth]{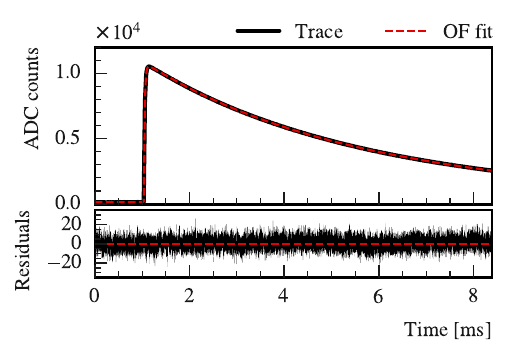}
\caption{\label{fig:of} Typical raw trace (black solid) and template scaled with the amplitude from an optimum filter-based fit (red dashed). The bottom panels shows the residuals, defined as the difference between raw trace and scaled template.}
\end{figure}

The OF algorithm requires the PSD of the noise $J(\nu)$ and the signal template $A(\nu)$, which are obtained via an iterative process.
Initially, a set of traces with similar height is selected and averaged to obtain a simple template of the signal.
This is then matched to the traces in the time domain through template fitting, and the resulting value of $\chi_\mathrm{TF}^2$ is employed as an additional selection.
In the absence of baseline data, the noise is computed by subtracting the template from the traces.
To eliminate artifacts arising from the steeply rising edge, the noise PSD is calculated using the average of the latter half of each trace, which corresponds to approximately \SI{4}{\ms} of the falling edge.
The obtained PSD is thus defined for frequency bins $\Delta f$ twice the size required for the acquired traces: the finer binning is recovered through linear interpolation.
This PSD is used to further fit the traces using the previously estimated template, and the difference between the fit template and the trace is then used as noise trace for the PSD calculation.
The final noise PSD is shown in \cref{fig:psd}, where an example of noise trace is shown as residuals of \cref{fig:of}.

The OF fit is applied to the traces using the simple template described above and the interpolated PSD.
The resulting OF amplitudes show two distinct populations corresponding to the {\kalpha } and {\kbeta } lines.
On average the former exhibits a reduced $\chi^2$ roughly \SI{2}{\percent} lower than the latter.
This difference can be attributed to the fact that the simple template was produced using signals from \kalpha, while slight variations in the signal shapes between the two energy regions are observed.
Distinct templates are thus defined for each population by averaging the traces chosen from the {\kalpha } and {\kbeta } lines.
The traces are then processed using both templates and the same PSD, resulting in two different filters $\mathrm{OF}_\alpha$ and $\mathrm{OF}_\beta$.
The maximum relative difference between the amplitudes yielded by them remains below \num{4e-5}, an effect that is neglected for the estimation of systematic uncertainties.
A similar difference is obtained when comparing the templates derived from traces selected using the loose and tight selections.
No trace showed a preference for a non-null time shift, excluding jitter effects of the order of the time bin of \SI{256}{\ns}.
The {\kalpha } and {\kbeta } templates obtained from the loose selection are used in the following. 
An example of a typical raw trace with the OF-matched signal is shown in \cref{fig:of}.

\begin{figure}[b]
\centering
\includegraphics[width=\columnwidth]{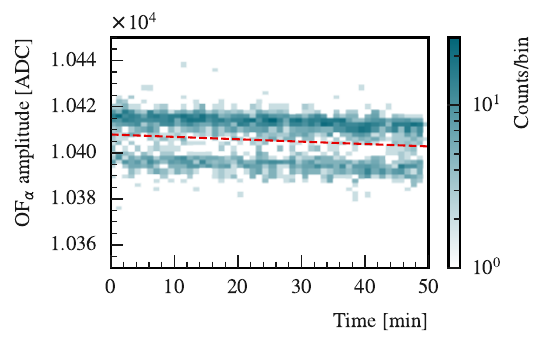}
\caption{\label{fig:cal1} Distribution of the fine-splitting structure of the {\kalpha } energy spectrum as a function of time.
The red dashed line indicates the time-dependent correction function.}
\end{figure}

Over the course of the calibration, the OF amplitudes steadily decreased by \SI{0.04}{\percent}, as shown in \cref{fig:cal1}.
While the exact nature of this effect remains uncertain, it is likely attributed to a temperature drift either of the detector itself or of the associated electronics outside the cryostat.
The amplitude decrease is well described by a first-order polynomial, which is fitted to both the $\mathrm{OF}_\alpha$ and $\mathrm{OF}_\beta$ amplitudes of the {\kalpha } line.
The obtained linear relations are used to scale the OF amplitudes and account for their time-dependence.
For the rest of this work, ``cOF'' amplitudes indicates the corrected amplitudes coming from $\mathrm{OF}_\alpha$ when considering {\kalpha } events, and from $\mathrm{OF}_\beta$ when considering {\kbeta } events.
Efforts were made to incorporate temperature information from the baseline mean by implementing a two-dimensional correction including both baseline mean and time, but with no improvement compared to the time-only correction.

\subsection{\label{sec:spectrum}Event spectrum and energy resolution}
The spectral shape of the characteristic X-rays from {\kalpha } and {\kbeta } lines of {\fe } was studied in detail in \cite{PhysRevA.56.4554,SAKURAI2003391}.
They are described as the superposition of Voigt distributions $\mathcal{V}\left(x\right)$, consisting of a Lorentzian with central energy $E_0$ and natural width $\Gamma$ which is convolved with a Gaussian detector response of variance $\sigma_E^2$.
This variance is correlated to the energy resolution of the detector, defined as the full-width half-maximum (FWHM) $\Delta E_\mathrm{FWHM} = 2\sqrt{2\ln{2}}\,\sigma_E$.
Each Voigt distribution is multiplied by an error function $\Phi\left(x\right)$ modeling the remaining contribution of the athermal phonons escaping into the substrate of the MMC before thermalizing, leading to a partial energy loss.
The probability density function (pdf) for the spectral lines is thus defined as
\begin{equation}
\begin{split}
    f^\mathrm{esc}_{\alpha/\beta}\left(E\,|\,\sigma_E,\varepsilon_\mathrm{esc}\right) = \sum_i^{\alpha/\beta}k_i&\mathcal{V}\left(E\,|\,E_0^{(i)}, \Gamma^{(i)}, \sigma_E\right)\\
    \times&\Phi\left(\varepsilon_\mathrm{esc}\left(E-E_0^{(i)}\right)\right),
\end{split}
\end{equation}
where $k_i$ is the relative intensity of the different contributions and $\varepsilon_\mathrm{esc}$ describes the impact of the escaping phonons.
The values describing the natural shape of the X-rays are taken from \cite{PhysRevA.56.4554}, where the {\kalpha } line is corrected after private communication with the author, as in \cite{krantz_thesis}.
\begin{table}
\begin{tabular}{c|c|c|c} 
 & Energy [eV] & Rel. intensity & Natural width [eV] \\
\hline \multirow{2}{*}{\centering \textbf{$\mathrm{K}_{\alpha 2}$}} & 5886.495 & 0.100 & 4.216 \\
& 5887.743 & 0.372 & 2.361 \\
\hline \multirow{6}{*}{\centering \textbf{$\mathrm{K}_{\alpha 1}$}} & 5894.829 & 0.068 & 4.499 \\
& 5896.532 & 0.096 & 2.663 \\
& 5897.867 & 0.264 & 2.043 \\
& 5898.853 & 0.790 & 1.715 \\
& 5899.417 & 0.007 & 0.969 \\
& 5902.712 & 0.0106 & 1.5528 \\
\hline \multirow{5}{*}{\centering \textbf{$\mathrm{K}_{\beta}$}} & 6477.73 & 0.077 & 13.22 \\
& 6486.31 & 0.109 & 9.40 \\
& 6488.83 & 0.176 & 2.81 \\
& 6490.06 & 0.397 & 1.81 \\
& 6490.89 & 0.608 & 1.83 \\
\end{tabular}
\caption{Central energies, natural widths and relative intensities used to describe the shape of the energy spectrum from the electron capture of {\fe }. The values come from \cite{PhysRevA.56.4554}, but with few changes for {\kalpha } following private communications with the authors, as in \cite{krantz_thesis}.}
\end{table}

\begin{figure}[t]
\centering
\includegraphics[width=\columnwidth]{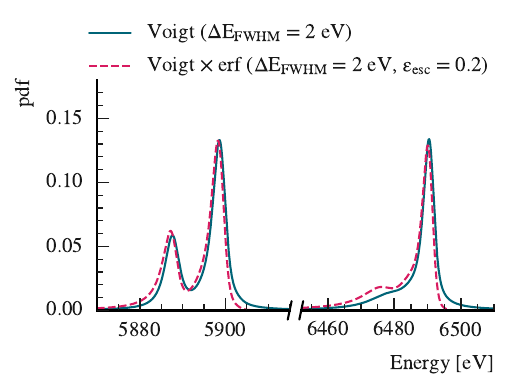}
\caption{\label{fig:models} Probability distribution function of {\kalpha } and {\kbeta } lines assuming an energy resolution of \SI{2}{\electronvolt}. Both the model including and excluding the impact of athermal phonons escaping the sensor are shown.}
\end{figure}

Both the scenarios including the escaping loss ($f^\mathrm{esc}_{\alpha/\beta}\left(E\,|\,\sigma_E,\varepsilon_\mathrm{esc}\right)$) and excluding it ($f_{\alpha/\beta}\left(E\,|\,\sigma_E\right)$) are considered in this work and their pdfs are plotted in \cref{fig:models}.
The calibration function correlating the cOF amplitude $A_\mathrm{cOF}$ to the energy is described by a second-order polynomial with no constant component:
\begin{equation}
    E = p_1\cdot A_{\mathrm{cOF}}^2 + p_2\cdot A_{\mathrm{cOF}},
\end{equation}
as suggested by a fit of the amplitude peak positions.

The $A_\mathrm{cOF}$ spectrum from the {\fe } calibration is fitted only for its {\kalpha } component.
This is accomplished through the $\chi^2_\alpha$ minimization for binned data, where the subscript refers to the spectrum component being considered.
This is done both for the model including the athermal phonon escape $a\cdot f^\mathrm{esc}_\alpha\left(E\,|\,\sigma_E,\varepsilon_\mathrm{esc}\right)$ and for the model excluding it $a\cdot f_\alpha\left(E\,|\,\sigma_E\right)$, where $a$ is the scaling factor.
The minimization returns the best estimators of the free parameters $a$, $\sigma_E$, $p_1$, $p_2$ and, when included, $\varepsilon_\mathrm{esc}$.
The $\chi^2_\alpha$ is minimized using the \textit{iminuit} software tool \cite{iminuit}.
Due to the multidimensional and stochastic nature of the problem, different bin sizes and initial parameters are considered.
This results in a distribution of best estimators for four different configurations, differentiated by the model used for the fit (including or excluding the athermal phonon escape) and by the thermal fluctuation selection (tight or loose).
The energy spectrum in the energy range of the {\kbeta } line is reconstructed for each set of estimated parameters under the assumption of non-varying energy resolution and athermal phonon escape contribution, when included.
The $\chi^2_\beta$ value estimated in the {\kbeta } region is used as an additional quality information about the performance of the individual fitting results.

\begin{figure}[t]
\centering
\includegraphics[width=\columnwidth]{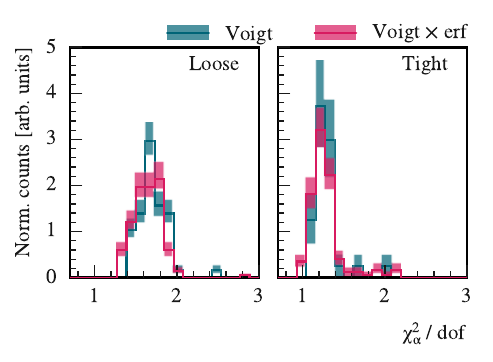}
\caption{\label{fig:chi2alpha} Distribution of the $\chi^2_\alpha$ for the loose (left panel) and tight (right panel) selections on the temperature fluctuations. Both model including the athermal phonon escape (magenta) and excluding it (dark blue) are shown.}
\end{figure}

The value of $\chi^2_\alpha$ indicates the goodness-of-fit of the considered model in the fit region, ie., \kalpha.
The distribution for the four different configurations is shown in \cref{fig:chi2alpha}, where a quality criterion on the goodness-of-fit in the {\kbeta } region is applied by requiring $\chi^2_\beta<1.2$.
The introduction of the factor modeling the escape of athermal phonons does not improve the quality of the fit, while a tighter selection on the temperature fluctuations clearly does.
Consequently, the performance of the detector is evaluated considering the dataset with the tight selection criterion applied and without including an escaping factor in the model.
This mitigates the temperature fluctuations that ultimately increase the measured energy resolution $\sigma_E$.

\begin{figure}[t]
\centering
\includegraphics[width=\columnwidth]{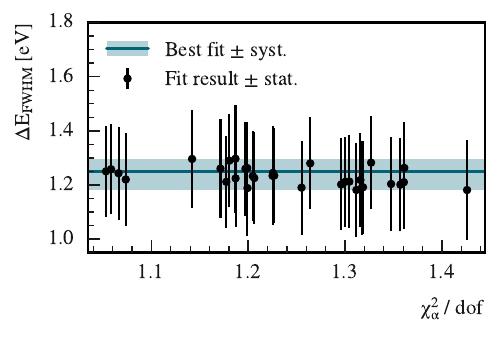}
\caption{\label{fig:fwhm_vs_chi2alpha} Energy resolution $\Delta E_\mathrm{FWHM}$ as from $\chi^2$ minimization and its statistical uncertainty as a function of the corresponding $\chi^2_\alpha$ (black circles). The systematic uncertainty is indicated by the blue shaded area. The best fit value with the lowest $\chi^2_\alpha$ is reported as the MMC energy resolution and it is indicated by the blue horizontal line.}
\end{figure}

\begin{figure}
\includegraphics{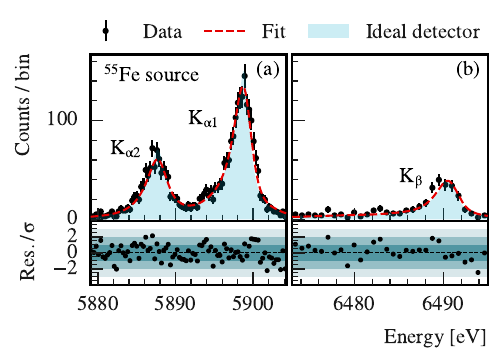}
\caption{\label{fig:fit} Measured spectrum of the (a) {\kalpha } and (b) {\kbeta } lines from the {\fe } calibration source. The dashed red line is the result from the fitting procedure with the lowest $\chi^2_\alpha$, while the filled area is the natural line shape assuming an ideal detector resolution. The residuals are shown in the bottom panels in units of standard deviation $\sigma$, were the 1-, 2-, and 3-sigma bands are shown as shaded areas.}
\end{figure}
The best-performing fit results are selected by requiring $\chi^2_\alpha<1.5$ and $\chi^2_\beta<1.2$.
This selection returns a distribution of energy resolutions which is shown in \cref{fig:fwhm_vs_chi2alpha} as a function of $\chi^2_\alpha$.
The value returning the minimum $\chi^2_\alpha$ is considered the energy resolution with the statistical uncertainty coming from the fit.
The systematic uncertainty is estimated by considering the highest and lowest value of the FWHM in the selected population.
The final value is $\Delta E_\mathrm{FWHM} = \left(\num{1.25(17)}\text{\scriptsize{(stat)}}^{+0.05}_{-0.07}\text{\scriptsize{(syst)}}\right)\unit{\electronvolt}$.
This result corresponds to an energy resolving power of \mbox{$E/\Delta E_\mathrm{FWHM}\simeq \num{4700(700)}$}.
The large statistical uncertainty arises from the small value of the energy resolution compared to the natural width of the $\mathrm{K}_{\alpha1}$ line of $\Delta E_{\alpha1}=2.47~\mathrm{eV}$~\cite{PhysRevA.56.4554}.
The fitted {\kalpha } and {\kbeta } spectra are shown in \cref{fig:fit} together with the their natural shapes, obtained by requiring $\Delta E_\mathrm{FWHM} = 0$.
When ignoring the impact of temperature fluctuations and including the escaping athermal phonon component, the energy resolution increases to $\Delta E_\mathrm{FWHM} = \left(\num{1.35(15)}\text{\scriptsize{(stat)}}^{+0.03}_{-0.09}\text{\scriptsize{(syst)}}\right)\unit{\electronvolt}$.
This is the most conservative scenario for the energy resolution, which should be considered to have this as upper value, although this scenario is disfavored by the fit results when comparing the models with and without athermal escaping contribution.
The energy resolution is compared to the baseline resolution of the OF, given by:
\begin{equation}
    \Delta E_\mathrm{FWHM}^\mathrm{OF} = 2\sqrt{2\ln{2}}\left(\sum_\nu\frac{|A(\nu)|^2}{J(\nu)}\right)^{-0.5},
\end{equation}
where the factor $|A(\nu)|^2/J(\nu)$ is known as noise-equivalent power \cite{Wilson:2022quj}.
This value corresponds to the expected resolution for a detector in thermal equilibrium and it represents the best resolution attainable with the OF given the noise condition of the detector.
The baseline resolution obtained using the information shown in \cref{fig:psd} corresponds to approximately \SI{1.1}{\electronvolt}, confirming the good performance of this analysis.
The resulting resolution can be compared to the measured thermal stability of $\Delta T/T\simeq\num{6.4e-4}$, dominated by the precision of the refrigerator thermometer.
At first order, the energy resolving power linearly depends on the thermal stability, meaning that the expected resolving power is $E/\Delta E_\mathrm{FWHM}\sim T/\Delta T\simeq\num{1600}$.
As the observed resolving power is approximately a factor 3 larger, this indicates that the sensor has a better temperature stability than what assessed for the refrigerator, indicating the need for an improved temperature monitoring for future measurements.

\section{\label{sec:conclusion}Conclusion}
MMCs stand out as one of the most promising technologies within the fast-growing field of cryogenic detectors, owing to their excellent energy resolution, large dynamic range and good linearity~\cite{Kempf:2018}.
This work presented the analysis of {\fe } calibration data acquired with the MMC detector discussed in~\cite{MMC_paper}, directly integrating the paramagnetic sensor into the dc-SQUID loop and improving the calorimeter design.
The amplitudes of the acquired pulses are extracted using an optimum filter-based algorithm and they are used to reconstruct the energy of the event.
After applying quality selections and a time-dependent correction, the obtained energy spectrum is fitted to the natural shape of the {\kalpha } spectral line convolved with a Gaussian energy resolution.
The calibration parameters obtained from the fit are used to reconstruct the {\kbeta } energy region of the spectrum and its goodness-of-fit is used to validate the results.
A systematic study of this fit procedure returns a best-fit resolution of $\smash{\Delta E_\mathrm{FWHM} = \left(\num{1.25(17)}\text{\scriptsize{(stat)}}^{+0.05}_{-0.07}\text{\scriptsize{(syst)}}\right)\unit{\electronvolt}}$.
To our knowledge, this results in the best energy resolving power achieved with an energy-dispersive detector in this energy range, at $E/\Delta E_\mathrm{FWHM}\simeq 4700$.
The analysis improves the resolution by \SI{30}{\percent} upon the previous template fitting-based analysis~\cite{krantz_thesis}.
Notably, the analysis also shows that the introduction of a factor to account for the escape of athermal phonons does not improve the final result.
However, the energy resolution deteriorates with a looser selection on the baseline mean, which serves as a proxy for the temperature of the detector.
This suggests that the energy resolution of the MMC is limited by the temperature fluctuations: the calorimeter could potentially reach sub-eV resolution with the implementation of a proper temperature stabilization.
In addition to presenting the world-leading energy resolution achieved by the X-ray MMC, this paper serves as the foundational work for the analysis and processing framework of the MMC-based \delight dark matter experiment.

\begin{acknowledgments}
It is a great pleasure to thank (in alphabetical order) Klaus Eitel, Christian Enss, Felix Kahlhoefer, Sebastian Lindemann, Marc Schumann, Kathrin Valerius, Matthew~Wilson and Alexander~Zaytsev for discussions,
and Klaus Eitel, Kathrin Valerius and Sebastian Lindemann for feedback and comments on earlier versions of the manuscript.
We further thank Matthew~Wilson and Alexander~Zaytsev for the help in the development of the software used for the processing of the MMC traces.
This work was supported by the Deutsche Forschungsgemeinschaft (DFG) through the Emmy Noether Grant No.~420484612, and by the Alexander von Humboldt Foundation.
\end{acknowledgments}

\bibliography{literature}

\end{document}